\documentclass[12pt]{article}
\usepackage[dvips]{graphicx}
\def \bsigma{\mbox{\boldmath $\sigma$}}

\def \bM{\mathbf{M}}
\def \bG{\mathbf{G}}

\def \bj{\mbox{\boldmath$j$}}
\def \br{\mbox{\boldmath$r$}}
\begin{document}
\begin{flushright}
GEF-TH-11/2006
\end{flushright}
\begin{center}
\noindent {\bf {The baryon octet magnetic moments to all orders in
flavor breaking; an application to the problem of the strangeness
in the nucleon}} \vskip 25pt G.Dillon and G.Morpurgo
\end{center}
Universit\`a di Genova and Istituto Nazionale di Fisica Nucleare, Sezione
di Genova.
\footnote{e-mail: dillon@ge.infn.it \hskip0.1cm ;\hskip0.2cm
morpurgo@ge.infn.it}

\vskip 20pt \noindent {\bf Abstract.} Using the general QCD parametrization (GP) 
we display the magnetic moments of the octet baryons including all flavor 
breaking terms to any order. The hierarchy of the GP parameters allows to  estimate 
a parameter $g_{0}$ related to the quark loops contribution of the
proton magnetic moment; its magnitude is predicted to be inside a
comparatively small interval including the value given recently by
Leinweber et al. from a lattice QCD calculation.
\baselineskip16pt \vskip15pt
 \noindent {\bf 1. Introduction}\vskip5pt
\baselineskip24pt The $s\bar{s}$ contribution $\mu_{p}^{s}=
-\frac{1}{3} G_M^s{(0)}$ to the proton magnetic moment (compare,
for the experimental situation, Refs.\cite{Sample,HGM}) has been
the object of many QCD calculations leading to a variety of
results. A list of references is found in the table 3 of the
review of Beck-McKeown \cite{Beck} and also in recent papers by
Leinweber et al. \cite{Lein}. Although, as we shall see, our
results are compatible with the most recent ones of Leinweber et
al. \cite{Lein}, it is known that QCD (either in the chiral or in
the lattice approaches) has produced in the course of the years a
variety of predictions for $G_M^s{(0)}$, differing in magnitude by
a factor
larger than 10 (and also differing in sign).\\
\indent To treat this problem we use here the general QCD
parametrization (GP) (\cite{mo89}-\cite{dm962}) of the magnetic
moments of the baryons, including (for the first time) terms of
all orders in flavor breaking. Exploiting a dynamical property of
the GP, its ``hierarchy'' of the parameters \cite{mo92}, we can
estimate the order of magnitude of a parameter $g_{0}$ related to
a part of the $s$ quark loops contribution of the proton magnetic
moment ($g_{0}$ is -3$O_{N}$ in the notation of Ref.\cite{Lein}).
The value of $g_{0}$ is predicted inside a rather narrow interval
that includes the value obtained by Leinweber et al. \cite{Lein}
by a lattice QCD calculation.\footnote{Recently \cite{dm06},
applying the GP to this problem, we related the key formula of
Beck-McKeown \cite{Beck} for $\mu_{p}^{s}\equiv -(1/3)G^{s}_{M}$,
to the quark loops corresponding to the $Z$-trace terms. Note: In
our notation the negative charge of $s$ is not included in the
definition of $G_M^s(0)$.}.

\vskip20pt \noindent {\bf 2. The general QCD parametrization of
the magnetic moments} \vskip5pt Consider the operator
\begin{equation}
    \bM  = (1/2)\int d^{3}\br\, (\br\times \bj (\br))
     \label{exM}
\end{equation}
where the current
\begin{equation}
    j_{\mu}(x)= e\bar{q}(x)Q \gamma_\mu q(x)
    \label{emc}
\end{equation}
is expressed in terms of the quark $(u,d,s)$ fields $q(x)$ and of
a charge-diagonal flavor matrix $Q$. The GP  parametrizes (compare
\cite{mo89} and \cite{dm96}-see there in particular
footnote 14) the expectation value of $\bM$ (Eq.(\ref{exM})) on the exact 
eigenstate $|\psi_B\rangle$ of the QCD Hamiltonian for the baryon
$B$ at rest as:
\begin{equation}
    \langle \psi_{B}|\bM | \psi_{B}\rangle =\langle \Phi_{B}|V^{\dagger}\bM V| \Phi_{B}\rangle
    =\langle W_{B}| \sum_{\nu} g_{\nu}\bG_{\nu}(\bsigma,f)| W_{B}\rangle
    \equiv\langle W_{B}|\bG(\bsigma,f)| W_{B}\rangle
    \label{VMV}
\end{equation}
The symbols in the above equations were used repeatedly in
previous work (\cite{mo89}-dm{962}), but, for convenience, we
recall part of them again: Thus $|\Phi_{B}\rangle$ is an
\textit{auxiliary} three quarks-no gluon state -having simple
properties (e.g. $L=0$); $V$ is a QCD unitary transformation
leading from the auxiliary state $|\Phi_{B}\rangle$ to the
\textit{exact} QCD eigenstate $|\psi_B\rangle$ ($|\psi_B \rangle$=
$V|\Phi_{B}\rangle$). In Eq.(\ref{VMV}) the space coordinates are
eliminated in the second step, as described in \cite{mo89};
the $\textbf{G}_{\nu}$'s are operators, linear in the matrix $Q$, depending 
only on the
spin $\bsigma$ and flavor $f$ of three quarks; $|W_{B}\rangle$ are the usual 
spin$(1/2)$-flavor octet states (with spin up); the $g_\nu$'s are  
parameters independent of  $Q$, the same for the magnetic moments 
($Q=Q^\gamma =Diag[2/3,-1/3.-1/3]$) and the Z-moments ($Q=Q^Z
\propto
Diag[1-(8/3)\sin^2\theta_W,-1+(4/3)\sin^2\theta_W,-1+(4/3)\sin^2\theta_W]$).

In the limit of exact $SU(3)$ ($m_u=m_d=m_s$) $\bG(\bsigma,f)$ is the sum of only three 
possible operators. Writing: $\bG_1=\sum_i Q_i \bsigma_i,\:
\bG_3=\sum_{i\ne k}Q_i \bsigma_k$, it is:
\begin{equation}
\bG(\bsigma,f) =g_{1}\bG_1 + g_{3}\bG_3 +
\hat{g}_{0}Tr[Q]\sum_i\bsigma_i
 \label{Gsym}
\end{equation}
In $\bG_1,\:\bG_3$
the index $i$ on $Q_i$ arises from 
the $\gamma$ (or $Z$) being attached to a quark line with that
index. The meaning of a quark line and its index in the exact
state $|\psi_B\rangle$ is illustrated in the fig.1 of
Ref.\cite{mo92}.\footnote{We recall that in the figure just cited
the boxes describe the effect of the transformation $V$ in
Eq.(\ref{VMV}) on the three quarks in $|\Phi_B\rangle$. A quark
line may zigzag inside the box as much as one likes, due to
emission and reabsorption of gluons; also each box contains quark
loops and gluon lines connected in
all possible ways consistent with the prescriptions of QCD.}\\
\indent The $Trace$ terms (e.g. the last term in Eq.(\ref{Gsym})
and others to appear in the following sections) can be due
only to quark loops; each loop is connected to the other quark 
lines by gluons; because the quark loops associated to $Trace$
terms including a charge must have on their contour an incoming
$\gamma$ (or $Z$), the number of gluons connecting these loops to
the other quark lines must be at least three, due to color
conservation and the Furry theorem (see, e.g., Sect.4 of
\cite{dmZS}). From Eq.(\ref{Gsym}), the contribution to the proton
magnetic moment from the $s$-quarks would be $(-1/3)\hat{g}_0$,
but, in the e.m. case, this $s$ quark
contribution is cancelled by those of the $u$ 
and $d$ quark loops, due to
\begin{equation}
Tr[Q^\gamma]=0 \label{trQ}
\end{equation}
\indent The symmetry breaking due to the heavier mass of $s$ (we work here 
assuming charge symmetry $m_u=m_d\equiv m$) is represented in 
the GP by terms containing an $s$ projector $P^s=Diag[0,0,1]$ or
the product of two such projectors with different indices
(products of more than two $P^s$ with different indices do not
contribute to matrix elements with octet states). Below we list
all possible $P^{s}$ structures.
\vskip 20pt \noindent{\bf 3. Flavor breaking in the GP}\\
\indent i){\it Terms with one $P^s$}

The terms with one $P^{s}$ to be considered in this subsection
were often indicated in the past as first order flavor breaking;
but it was \textit{underlined} since the beginning \cite{mo89})
that, because $P_{i}^{s}\equiv (P_{i}^{s})^{n}$ for any $n$, such
terms contained also contributions of higher order
in flavor breaking ($\Delta m/m_{s}\equiv (m_{s}-m)/m_{s}$).\\
\indent There are five operators with one indexed $P^s$ (see
\cite{mo89,dm96}): They are: ${\bf G}_2=\sum_i Q_i P_i^s
\bsigma_i$ ; ${\bf G}_4=\sum_{i\ne k} Q_i P_i^s \bsigma_k$ ;
 ${\bf G}_5=\sum_{i\ne k} Q_k P_i^s \bsigma_i$ ; ${\bf
G}_6=\sum_{i\ne k} Q_i P_k^s \bsigma_i$ ; ${\bf G}_7 =\sum_{i\ne
j\ne k} Q_i P_j^s \bsigma_k$. Of course they act only on the
strange baryons. There is also a $Trace$ term containing one
$P^{s}$ (that acts, clearly, also on $p$ and $n$):
\begin{equation}
\bG_0=g_0Tr[QP^s]\sum_i\bsigma_i
\label{G0}
\end{equation}
The magnetic moments $\mu_B$ for the octet baryons $B$ are:
\begin{equation}
\mu_B=\langle \psi_B|(\bM)_z |\psi_B\rangle=\langle W_B|
\sum_{\nu=0}^7\: g_\nu (\bG_\nu)_z|W_B\rangle \label{mmPs1}
\end{equation}
 In Eqs.(\ref{G0},\ref{mmPs1}) $g_0$ is associated to the 
flavor breaking $s$-quark contribution to the proton moment, so
that
\begin{equation}
G_M^s(0)=\hat{g}_{0}+g_{0}    \label{g0t}
\end{equation}
(multiplied by $-1/3$) is the whole $s$-contribution to $\mu_p$
that one is measuring in the $\gamma$-Z interference experiments
\cite{Sample,HGM}.

At this stage one has 8 parameters and 8 data for the octet (7 magnetic moments 
plus the $\Sigma\Lambda$ transition moment) but the expectation values of the $\bG_\nu$'s 
($\nu=0,\ldots 7$) in the $|W_B\rangle$'s are not independent (see
the identity in Eq.(8) of \cite{dm962}). Thus \cite{dm06} it is
impossible to derive the strangeness content of the nucleon only
from the octet-baryons magnetic moments without exploiting some
additional property of QCD.

\indent We now list the flavor breaking structures with 2 (or
more) $P^{s}$.
\vskip 10pt \indent ii) {\it Terms with two (or more) $P^s$}

To proceed we consider the following flavor breaking structures
with two $P^{s}$ (they act only on $\Xi^0$ and $\Xi^-$):
\begin{equation}
\begin{array}{lcl}
{\bf G}_{7b}=\sum_{i\ne k} Q_iP_i^sP_k^s \bsigma_i & {\bf G}_{7c}=\sum_{i\ne k} 
Q_i P_i^s P_k^s\bsigma_k & \\ \\
 {\bf G}_8=\sum_{i\ne j\ne k (j>k)} Q_i P_j^s P_k^s\bsigma_i&
 {\bf G}_9=\sum_{i\ne j\ne k} Q_i P_k^s P^s_j\bsigma_k & {\bf G}_{10}
=\sum_{i\ne j\ne k} Q_i P_i^s P^s_k\bsigma_j
\end{array}
\label{Ginu}
\end{equation}
\indent We have also 4 $Trace$ structures of higher order:
\begin{equation}
\begin{array}{lr}
{\bf G}_0^{a}=Tr[QP^s]\sum_{i} P_i^s \bsigma_i & {\bf G}_0^{b}=Tr[QP^s]\sum_{i\ne 
k} P_i^s \bsigma_k \\ \\
 {\bf G}_0^{c}=Tr[QP^s]\sum_{i\ne k} P_i^s P_k^s\bsigma_k \quad\quad&
 {\bf G}_0^{d}=Tr[QP^s]\sum_{i\ne j\ne k (i>j)} P_i^s P^s_j\bsigma_k
\end{array}
\label{Ginu0}
\end{equation}
\vskip 10pt
Using Eq.(\ref{VMV}) 
with $Q=Q^\gamma$, one gets the magnetic moments as 
linear combinations of the parameters $g_\nu$'s. Because the 9
parameters multiplying the structures in
Eqs.(\ref{Ginu},\ref{Ginu0}) intervene only in a few combinations
\footnote{On using $Q_{i}P_{i}^{s}= -(1/3)P_{i}^{s}$ and
$Tr(QP^{s})= -(1/3)$ many operators in (\ref{Ginu},\ref{Ginu0})
reduce to the same, or to some already considered (e.g. ${\bf
G}_{0}^{a}\equiv{\bf G}_{2}$). One may ask why these structures
are listed separately; the reason is that they correspond to
different QCD diagrams; only displaying them separately, the order
of magnitude of their parameters can be estimated by the
``hierarchy"- see Sect.4.}, it is convenient to define:
\begin{equation}
\begin{array}{l}
g_{\Xi}=\frac{2}{9}(-2g_{7b}-2g_{7c} -g_8+4g_9+g_{10})\\
g_0^\Lambda=g_0+g_0^a\\
g_0^\Sigma=g_0-g_0^a/3+4g_0^b/3\\
g_0^\Xi=g_0+4g_0^a/3+2g_0^b/3 +4g_0^c/3 -g_{0}^{d}/3
\end{array}
\label{gxi}
\end{equation}
We get finally (compare Eq.(\ref{mfg})) the magnetic moments to
all orders in flavor breaking (in Eq.(\ref{mfg}) the baryon symbol
indicates the magnetic moment). The contributions of the $1P^s$
terms (see, e.g. \cite{dm962}) are obtained by omitting in
(\ref{mfg}) $g_8, g_9$ and the terms with $g_\Xi, g_0^\Lambda,
g_0^\Sigma, g_0^\Xi$ defined in the Eq.(\ref{gxi}) above:
\begin{equation}
\begin{array}{l}
\label{mfg}
p= g_1-g_0/3\\
n=-(2/3)(g_1-g_3)-g_0/3 \\
\Lambda =-(1/3)(g_1-g_3+g_2-g_5)-g_0^\Lambda/3\\
\Sigma^+=g_1+(1/9)(g_2-4g_4-4g_5+8g_6+8g_7)-g_0^\Sigma/3\\
\Sigma^-=-(1/3)(g_1+2g_3)+(1/9)(g_2-4
g_4+2g_5-4g_6-4g_7)-g_0^\Sigma/3\\
\Xi^0=-(2/3)(g_1-g_3)+(2/9)(-2g_2-g_4+2g_5-4g_6+5g_7)+g_{\Xi}-g_0^\Xi/3\\
\Xi^-=-(1/3)(g_1+2g_3)+(2/9)(-2g_2-g_4-4g_5-g_6-g_7+ \frac{3}{2}g_8-6g_9 
)+g_{\Xi}-g_0^\Xi/3\\
\mu(\Sigma\Lambda)=-\frac{1}{\sqrt{3}}\Big(g_1-g_3+g_6-g_7 \Bigl)
\end{array}
\end{equation}
\indent From Eqs.(\ref{gxi},\ref{mfg}) one gets Eq.(\ref{okubo}),
also QCD exact to all orders in flavor breaking.
 Note that $\mu(\Sigma\Lambda)$ can
have an imaginary part \cite{mo89}; in principle this might
contribute, via other terms, to the
$\Sigma^0\rightarrow\gamma\Lambda$ decay \textit{rate}.
But, because all \textit{open} $\Sigma^0\rightarrow\Lambda$ channels 
are $\gamma$ channels, such terms would imply additional
$\gamma$'s and are therefore negligible.
\begin{equation}
\sqrt{3}\ 
\mu(\Sigma\Lambda)=n-(3/2)\Lambda-(\Sigma^++\Sigma^-)/4+\Xi^0-g_\Xi+(4/9)g_0^c
-(1/9)g_0^d \label{okubo}
\end{equation}
The Eq.(\ref{okubo}) (barring the higher order parameters 
$g_\Xi$, $g_0^c$, $g_0^d$) is the Okubo relation \cite{okubo}.

From the previous formulas one can calculate the contribution of
each specific flavor $f=(u,d,s)$ to the baryon magnetic moments
and compare the resulting equations with Eqs.(4,5) of
Ref.\cite{Lein}, obtained exactly from QCD by Leinweber et al.
using only $u,d$ charge symmetry.\footnote{Charge symmetry for $u$
and $d$ has always been part of our general QCD parametrization,
although, of course, in some cases -e.g. for the explanation of
the extraordinary level of precision of the Coleman-Glashow e.m.
baryon mass formula \cite{dm2000E}- its applicability was examined
in detail (see Sect.4 of \cite{dm2000E}).} The contribution of the
flavor $f$ to the magnetic moment of B (except for the loops and
to be still multiplied by the charges) is indicated as $f^{B}$
similarly to Ref.\cite{Lein}. In the GP treatment $u^{p},u^{n}$, etc. are calculated as follows: 
Replace the matrix $Q$ in Eq.(\ref{emc}) with $P^f$. We then obtain for 
$u^{p},u^{n}$, etc. the results:\footnote{To obtain (referring to the $u$ quarks)
 the full contribution $\mu^{u}_{B}$ to the 
magnetic moment of baryon $B$, one must add to $u^{B}$ the loop
contributions -for the proton simply $\hat{g}_{0}$- and multiply
the result by the $u$-charge $Q_{u}$ (for $p$ and $n$ see the
Eqs.(20,21) of Ref.\cite{dm06})}
\begin{equation}
\begin{array}{l}
u^{p}= (4/3)g_1+(2/3)g_3 \\
u^{n}= (-1/3)g_1+(4/3)g_3 \\
u^{\Sigma^+}=(4/3)g_1+(2/3)g_3+(-2/3)g_5+(4/3)g_6+(4/3)g_7 \\
u^{\Xi^0}=(-1/3)g_1+(4/3)g_3+(4/3)g_5+(-2/3)g_6+(4/3)g_7+(-1/3)g_8
+(4/3)g_9 \label{uB}
\end{array}
\end{equation}
We easily find using Eqs.(\ref{mfg},\ref{uB}) the relations:
\begin{equation}
u^{\Sigma^+}=\Sigma^+ - \Sigma^- \qquad ; \qquad
u^{\Xi^0}=\Xi^0 - \Xi^-
\label{xisigma}
\end{equation}
\begin{equation}
u^p = 2p+n+g_0 \quad ; \quad u^n= p+2n+g_0
\label{LT}
\end{equation}
that are exact consequences of QCD (plus $u,d$ charge symmetry)
since the GP displays and takes into account
 all the  
operators $\bG_\nu$ compatible with QCD. Because of the entirely
different derivation, it is expected and satisfactory that
Eqs.(\ref{xisigma},\ref{LT}) lead to the same equations and
therefore to the same constraint (fig.2 of \cite{Lein}) between
the ratios $u^p/u^\Sigma$ and $u^n/u^\Xi$ as those of \cite{Lein}.
This is seen by inserting in Eqs.(\ref{LT}) the expression
(\ref{notation}) for $g_0$. To summarize: Because
Eqs.(\ref{xisigma},\ref{LT}) follow exactly from QCD plus charge
symmetry, the general parametrization developed above -that
includes exactly \textit{all possible terms consistent with QCD},
plus charge symmetry- must lead necessarily to the above
equations. Note, by the way, that some papers (e.g. \cite{cloet})
misinterpreted the GP as some model
``a little more sophisticated than the simplest constituent quark model".\\
\indent The connection of our $g_{0},\hat{g}_{0}$
(Eqs.(\ref{Gsym},\ref{G0},\ref{g0t})) with the symbols in
\cite{Lein} is:
\begin{eqnarray}
\label{notation}
g_0=-3O_N= -G_M^s(0)(1-^l\!R^s_d)/ ^lR^s_d\\
^lR^s_d= 1+g_0/\hat{g}_0
\end{eqnarray}

\vskip20pt \noindent {\bf 4. The hierarchy of the coefficients in
the general QCD parametrization}\vskip5pt \indent Everything, so
far, is an exact consequence of QCD (plus $u,d$ charge symmetry).
Now we turn to the determination of $g_0$. As already noted
\cite{dm06} its value cannot be determined from the magnetic
moments only. One has to exploit a dynamical property of QCD,
``the hierarchy of the parameters", as anticipated at the end of
Sect.1. The ``hierarchy" is an empirical property, noted long ago
\cite{mo92} and used in past work (e.g.\cite{moPRL,dm2000E});
below we recall only its main points. Although the hierarchy was
discussed in detail in the references cited above, for convenience
we will recall briefly (at the end of this section) how it works
in the first case to which it was
applied (the extension of the Gell Mann-Okubo mass formula to second order flavor breaking).\\
\indent It is a fact that the coefficients $g_{\nu}$ of the
various structures (the ``parameters") are seen to decrease with
increasing complexity of the structure. This was first seen in the
GP of the \textbf{8+10} baryon masses, but applies generally. It
appears from the previous analysis that each baryon property (e.g.
the magnetic moments) is QCD parametrized as a sum of structures
each being a sum of terms having a maximum of three different
indices (as an example, see the structure ${\bf G}_8=\sum_{i\ne
j\ne k (j>k)} Q_i P_j^s P_k^s\bsigma_i$); these indices are
appended to a quark charge, or to a $P^{s}$ or to a $\bsigma$ in
the term under consideration. If the rules of Ref.\cite{mo92} are
adopted to normalize univocally the sums, the coefficients of the
various structures - that is the parameters $g_{\nu}$ - decrease
in a well defined way with the number of different indices present
in the terms being summed and, of course, also with the number of
$P^{s}$ (flavor breaking) factors in the structure. Each factor
$P^s_i$ produces a reduction \footnote{Of course the reduction,
for the magnetic moments, is relative to $g_1\approx 2.8$.}
$\approx 1/3$ and each pair of indices produces a reduction -as
discussed in \cite{dm96}- by a factor from 0.2 to 0.37 (see the
footnote 16 of \cite{dm96}). This ``pair of indices" reduction is
due to the exchange of a gluon between two quark lines and applies
to baryons or mesons composed of quarks
$u,d,s$.\footnote{Incidentally these reduction factors explain why
the naive NRQM, based essentially on terms with few indices, works
reasonably; to understand this fact on the basis of QCD was
precisely the aim of the GP (see also \cite{mosta}).} In the
calculations performed so far (e.g.\cite{moPRL,dm2000E,dm99PL})
this hierarchy of the coefficients was used to show why certain
formulas (the Gell Mann-Okubo baryon mass formula, or the
Coleman-Glashow or the Gal-Scheck formulas) work so well. Observe
that to do this -that is to show that the terms neglected were
indeed negligible- any choice of the ``pair of indices" reduction
factor given above (from 0.2 to 0.37) was equivalent; we selected
$0.3$ or $0.33$ in the above evaluations for convenience (the same
factor as that due to $P^{s}$); however, since below we are going
to use the hierarchy to estimate the order of magnitude of
$g_{0}$, we will consider the whole interval 0.2 to 0.37 mentioned
above.\\
\indent To exemplify the above presentation we will summarize the
main points of the first application of the GP \cite{moPRL} to the
baryon masses. In the general parametrization the masses of the
eight lowest octet + decuplet baryons can be expressed in terms of
8 parameters (the e.m. mass differences are neglected). The known
values of the masses allow to calculate all the above parameters;
one can check that the parameters multiplying the two more
complicated structures (sums over 3 indices-containing 2 or 3
$P^{s}$) are negligible. Neglecting them one finds a new relation
between the masses which generalizes the Gell Mann-Okubo formula
including to a very good approximation the contribution of the
$2P^{s}$ terms.

\vskip20pt \noindent {\bf 5. Estimating $g_0$}\vskip5pt \indent
For the baryon octet magnetic moments at the $1P^s$ level it has
been seen repeatedly that the hierarchy for the $g_\nu$
($\nu=1...7$) applies fairly well except for the ratio $p/n$. It
is well known that for the ratio $p/n$ the hierarchy does not
work; indeed the extraordinary smallness of $(p/n +3/2)$ is
accidental; therefore one cannot determine the order of magnitude
of $g_0$ from the $p,n$ system alone. (For this compare
\cite{mo92} and the second paper in \cite{dm96}; see also
\cite{th}).\footnote{Actually the $M1$ transition $\Delta \to
p+\gamma$ allows to understand, via the hierarchy applied to the
whole octet-decuplet system, the smallness of $g_3$ ; compare the
Sects. V, VI -see the Eq.(22)- in \cite{dm96}.}

Leaving aside $g_3$, there is however a way to estimate the order
of magnitude of $g_0$ from Eqs.(\ref{mfg}); indeed from them one
derives the relation (\ref{g0}) (again, of course, an exact QCD
equation):
\begin{equation}
g_0=-\frac{5}{4}p-\frac{7}{4}n+\frac{\sqrt{3}}{2} \mu(\Sigma\Lambda) 
+\frac{1}{2}(\Sigma^+-\Sigma^-)+\frac{1}{4}(\Xi^0-\Xi^-)-\frac{3}{2}g_7+
\frac{1}{12}g_8- \frac{1}{3}g_9 \label{g0}
\end{equation}
Inserting experimental values in Eq.(\ref{g0}) 
one obtains (in nuclear magnetons):
\begin{equation}
g_0=0.13\pm 0.07-\frac{3}{2}g_7+\frac{1}{12}g_8-\frac{1}{3}g_9
\label{g0num}
\end{equation}
where the error comes essentially from $|\mu(\Sigma\Lambda)| =1.61\pm 0.08$ (
the sign of $ \mu(\Sigma\Lambda)$ is negative, as can be seen from 
Eq.(\ref{okubo})).\footnote{Neglecting $g_8$ and $g_9$, one gets from (\ref{g0num}) 
$\tilde{g}_7\equiv g_7+(2/3)g_0 \simeq 0.09\pm0.04$ instead of $\tilde{g}_7 \simeq 0.16$, the value given 
in the 2nd paper in \cite{dm96}. This is because, to determine
$g_{7}$, we now used $\mu(\Sigma\Lambda)= -1.61\pm0.08$ deduced
from its experimental value $\mid \mu(\Sigma\Lambda)\mid =
1.61\pm0.08$. (Instead in \cite{mo89},\cite{dm96} the $1P^{s}$
calculation was used
 giving $\mu(\Sigma\Lambda)=-1.48\pm 0.04 \mu_{N}$). A new measurement of
 $\mu(\Sigma\Lambda)$
 would be of interest; the present value is the result of a Primakoff effect determination
 \cite{sigmalam} dating back to 1986.}\\
\indent In (\ref{g0num}) $g_8$ and $g_9$ are parameters associated to a structure with three indices 
plus two $P^s$ operators; they are both expected to be of order  
$0.03$. Thus $\mid(g_{8}/12)-(g_{9}/3) \mid \leq 0.01$. So the
contributions of $g_8$ and $g_9$ can only affect negligibly the
already large uncertainty in the first term
of Eq.(\ref{g0num}). Instead, the parameter $g_7$ is associated to a structure with three 
indices plus one $P^s$; its contribution cannot be neglected.
 Depending on the choice of the gluon exchange
reduction factors in the hierarchy (from 0.2 to 0.37), one 
expects
\begin{equation}
     0.036\leq |g_7| \leq0.13
      \label{g7}
\end{equation}
We thus obtain from Eq.(\ref{g0num}), if $g_{7}$ is negative:
$\:0.18\pm0.07\leq g_{0} \leq 0.33\pm0.07$ and, if $g_{7}$ is
positive: $\:-0.07\pm0.07\leq g_{0} \leq +0.08\pm0.07$.
Altogether:
\begin{equation}
-0.07\pm0.07\leq g_0\leq 0.33\pm0.07 \label{g0inter}
\end{equation}
This interval includes $g_0=0.28\pm0.07$, the value that can be
extracted from Fig.2 and Eqs.(11) of Ref.\cite{Lein}. As a matter
of fact the environment dependence of the $u$-quark magnetic
moment as reflected by $u^n/u^{\Xi^0}$, discussed in \cite{Lein}
and re-derived above using the GP with charge symmetry -
Eqs.(\ref{xisigma},\ref{LT})- suggests to stay in the region of
positive $g_{0}$, so that the interval to be considered reduces
to:
\begin{equation}
0\leq g_0\leq 0.33\pm0.07 \label{g0intershort}
\end{equation}
\indent Because $u^n/u^{\Xi^0}=(g_0+2n+p)/(\Xi^0-\Xi^-)$ its value
is $\simeq 1.2$ for $g_0=0.33$ and $\simeq1.7$ for $g_0=0$. As to
$u^p/u^{\Sigma^+}=(g_0+2p+n)/(\Sigma^+-\Sigma^-)$ its values are
$\simeq1.1$ for $g_0=0.33$ and  $\simeq 1.0$ for $g_0=0$.\\

\vskip20pt \noindent {\bf 6. Conclusion} \vskip5pt The main point
of this paper has been to show that in $G^s_M(0)$
=$g_0$+$\hat{g}_0$ one can evaluate $g_0$ by performing a complete
general QCD parametrization of the baryon octet magnetic moments
and exploiting the GP ``hierarchy of parameters". The point of
interest is that by the simple general QCD parametrization one
obtains a value (compare Eq.(\ref{g0intershort})) compatible with
the value ($0.28\pm0.07$) obtained in Ref.\cite{Lein}. \\
\indent Of course $G_M^s(0)$ is the sum of $g_0$ and $\hat{g}_0$
(Eq.(\ref{g0t})), where $\hat{g}_0$ is related to $^lR^s_d$ of
Ref.\cite{Lein} by Eq.(18)[$^lR^s_d= 1+g_0/\hat{g}_0$]. Even if
$^lR^s_d$ had a value two times larger than that
 $0.139\pm 0.042$ of Ref.\cite{Lein}-as it seems possible-
 the value of $\mu^p_s$ would stay in the region of very small values
(that is at the limit of the present experimental possibilities)
recently indicated by HAPPEX (compare the HAPPEX JLAB report 2006
in Ref.\cite{HGM}) and its experimental determination (especially
at $Q^2=0$) would remain extremely difficult.\\
\indent [Note added, march 1, 2007] Recently the results of the
HAPPEX collaboration have been published \cite{HAPPEX} confirming
the above indications. While of course we refer to \cite{HAPPEX}
for any detail, we note only that the authors state that the
experiments will continue at values of $Q^2$ higher than the
present one $\approx 0.1\: GeV^2$. At such low values of $Q^2$ the
results (in the words of the authors) ``leave little room for
observable nucleon strangeness dynamics".

\end{document}